# Heterogeneous Unmanned Aerial Vehicles Cooperative Search Approach for Complex Environments

Zhenchang Liu,[1]  Mingrui Hao[2*]

*Science and Technology on Complex System Control and Intelligent Agent Cooperation Laboratory, Beijing, 100074, People's Republic of China*

**Abstract**

This paper studies a heterogeneous Unmanned Aerial Vehicles (UAVs) cooperative search approach suitable for complex environments. In the application, a fixed-wing UAV drops rotor UAVs to deploy the cluster rapidly. Meanwhile, the fixed-wing UAV works as a communication relay node to improve the search performance of the cluster further. The distributed model predictive control and genetic algorithms are adopted to make online intelligent decisions on UAVs' search directions. On this basis, a jump grid decision method is proposed to satisfy the maneuverability constraints of UAVs, a parameter dynamic selection method is developed to make search decisions more responsive to task requirements, and a search information transmission method with low bandwidth is designed. This approach can enable UAVs to discover targets quickly, cope with various constraints and unexpected situations, and make adaptive decisions, significantly improving the robustness of search tasks in complex, dynamic, and unknown environments. The proposed approach is tested with several search scenarios, and simulation results show that the cooperative search performance of heterogeneous UAVs is significantly improved compared to homogeneous UAVs.

## I. Introduction

Cooperative search approach has essential application value in civil and military fields [1]. It can be used for wild rescue [2], environmental monitoring [3], target reconnaissance [4], and other tasks. The core function of this approach is to help UAVs make decisions of search direction online in a broad unknown environment so that UAVs can quickly find targets and provide initial information for subsequent allocation [5] or monitoring [6] algorithms. In order to improve the search efficiency of UAVs, it is necessary to share detection information during the search process so that individuals within the cluster can avoid repeated searches for the same area. With the development of communication technology, a cluster's network can realize the functions of no central node, dynamic topology, and multi-hop transmission [7]. UAVs can share detection information and realize distributed autonomous decisions through an Ad-Hoc network [8].

---

[1] Senior Engineer.

[2] Researcher, Corresponding author.

In cooperative search tasks, due to the dynamic characteristics of the environment, clusters may face unexpected situations, such as unknown threats and individual damage. Hence, UAVs need to make online dynamic decisions. When UAVs make decisions, it is necessary to comprehensively consider the benefits (target existence probability, environmental uncertainty, etc.) and constraints (maneuverability constraints, communication constraints, space constraints, etc.) so that UAVs can search the task area safely and efficiently. In the state of arts, most of the work studied the cooperative strategy of homogeneous UAVs, considering only a few constraints. They did not achieve the ability of UAVs to respond to dynamic environments and make adaptive decisions.

This paper uses the fixed-wing UAV to launch rotor UAVs for the search task. After being launched, rotor UAVs enter the low-altitude space to search for targets, while the fixed-wing UAV acts as a communication relay node and hovers over the search area. Fixed-wing UAVs and rotor UAVs form a heterogeneous cluster to carry out the search task jointly. The distributed model predictive control (MPC) is used to calculate the search direction of UAVs timely. In this approach, the map is rasterized to make UAVs move in the discrete grid and improve the speed of online decisions. Moreover, a decision method in discontinuous grids is designed so that decision results can meet the maneuverability constraints of the UAV. The program parameters are dynamically selected to enable UAVs to adjust the decision tendency and flight speed according to environments. The distributed communication mechanism is adopted to enable UAVs to transmit search information under communication constraints. Compared to existing works, this work fully utilizes the differential advantages of heterogeneous UAVs in flight altitude and speed. It provides comprehensive solutions for complex issues in dynamic task scenarios. The main contributions of the research are as follows:

(1) A jump grid decision method is proposed to address the flight capabilities differences of heterogeneous UAVs, transforming the search decision problem into a single variable discrete optimization problem, ensuring the online decision-making speed, and enabling the decision results to meet maneuverability constraints.

(2) A parameters dynamic selection method is proposed to address the decision adaptability of UAVs at different stages of tasks. The cooperative search decision is achieved with adaptive capabilities by adjusting objective function weights, prediction sequence length, and jump grid value online.

(3) A search information transmission method for heterogeneous UAVs is proposed to address communication constraints in complex environments. The communication frequency and cooperative search efficiency are significantly improved by introducing a relay node.

The remainder of this paper is organized as follows. Section II discusses existing works. Section III introduces a cooperative search method based on distributed MPC. Section IV proposes the jump grid decision method. Section V proposes a parameter dynamic selection method. Section VI proposes a search information transmission method between UAVs. Section VII validates the proposed methods through simulation, while Section VIII summarizes the work. The program in simulation is open-sourced, and details are included in Section IX

## II. Related Work

The existing cooperative search methods mainly include geometric-based methods [4][9][10], heuristic methods [11][12][13][14][15], model predictive control (MPC) methods [16][17][18][19][20][21], reinforcement learning methods [22][23], etc. Among them, the pre-planned or fixed-pattern search strategies are not suitable for dynamic environments, centralized decision-making requires good network connectivity, and distributed decision-making has low requirements for communication topology and good robustness. The distributed MPC method rasterizes the map and describes the prior information in the grid through numerical values [16]. Based on factors such as the target existence probability, and environmental uncertainty, UAVs predict the search revenue after multi-step movement in different grids and make receding horizon decisions on the direction of movement through optimization solutions [18][19]. The distributed MPC method has the advantages of a multi-factor schedule and suitability for complex task environments. It has received widespread attention and application in the field of cooperative search.

When using the MPC to make search decisions for heterogeneous UAVs, it is necessary to comprehensively consider complex issues such as maneuverability constraints, decision adaptability, and communication constraints. References [17][18][19] took the heading angle as the decision object. They defined the variation amount of the heading angle available for decision as $-45°$, $0°$, $45°$, and the UAV can only move between adjacent grids. This method simplifies the search decision problem into a single variable discrete optimization problem, significantly reducing the online calculation load. However, UAVs with low maneuverability may need help to complete the $\pm 45°$ heading angle changes within two grid flight spaces, resulting in decisions that cannot be executed. If the grid size is increased to meet the maneuverability constraints of UAVs, it will decrease decision-making precision. Di et al. [20] took UAVs' speed and angular velocity as decision objects, transforming the search decision problem into a continuous optimization problem with multiple variables. Although this method considers the maneuverability constraints of UAVs, compared to the single variable discrete optimization problem, the computational complexity is significantly improved. In [13][14], based on the UAV's movement distance and heading angles range within each decision, all grids satisfying maneuverability constraints were selected as decision objects. Although this method also transforms the search decision problem into a single variable discrete optimization problem, the variable space will increase when the grid size is small, resulting in a large computation cost.

Regarding decision adaptability, Mou et al. [2] used the adaptive parameter processor to set the weight ratio in the objective function, enabling the UAVs to achieve the maximum search revenue in a limited time. However, this method used a traversal method to calculate the weight ratio, which requires significant computation. Liu et al. [21] designed an expert system to adjust the search parameters online, enabling UAVs to dynamically change their decision tendencies in cooperative search tasks. However, this method's parameter adjustment is discontinuous and unsuitable for UAVs in continuous maneuvering processes such as threat avoidance.

Regarding communication constraints, [19][20][24][25] adopted methods such as minimum spanning trees to ensure the connectivity of information transmission by controlling the topology shape of the cluster and the distance between UAVs. References [28][25][26] designed search information fusion methods to achieve the ideal search performance of UAVs under conditions such as communication interruption and data loss. However, the above studies have focused on homogeneous clusters and did not consider how to utilize individuals' height and speed differences in heterogeneous clusters to solve communication constraints.

In cooperative search tasks in three-dimensional space and multimedia environment, existing research used various unmanned vehicles such as unmanned aerial vehicles (UAVs), unmanned ground vehicles (UGVs), unmanned surface vehicles (USVs), and unmanned underwater vehicles (UUVs) to form heterogeneous clusters [27][28]. Kashino et al. [29] used UAVs and UGVs for the cooperative search of moving targets and confirmed the targets using UGVs after UAVs discovered them. This study used an iso-probability curve to design a search trajectory (spiral trajectory) based on the target's last known position, ensuring that the vehicle's detection area covers the target's moving range. Ke et al. [30] studied a cooperative search method for the air-sea heterogeneous system, which solved the problem of cross-medium transmission of information between different individuals through trajectory planning. However, the trajectory design in the above studies is relatively simple without considering large-scale clusters' collision and threat avoidance functions. The planning results cannot simultaneously meet multiple search requirements, such as target existence probability and environmental uncertainty, making it unsuitable for use in complex and dynamic task environments.

In response to the above issues, this paper studies a heterogeneous UAVs cooperative search approach based on a distributed MPC framework. It can enable heterogeneous UAVs to discover targets quickly, respond to unexpected situations in dynamic environments, and adapt to various constraints. The differences between the proposed solution and existing studies are summarized in Table 1.

Table. 1 comparison between the proposed solution with existing works

| Work | Path Planning Method | Cluster type | Maneu-const | Com-const | Obstacle avoidance | Dynamic threats | Adaptive decision |
|------|---------------------|--------------|-------------|-----------|--------------------|-----------------| ------------------|
| [4] | Coverage path planning | homo | ✓ | | | | |
| [10] | Coverage path planning | homo | ✓ | | ✓ | | |
| [12] | Ant colony optimization | homo | | ✓ | ✓ | | |
| [13][14] | Ant colony optimization | homo | ✓ | ✓ | ✓ | | |
| [15] | Pigeon-inspired optimization | homo | ✓ | | ✓ | | |
| [18][19] | Model predictive control | homo | | ✓ | | | |
| [20] | Model predictive control | homo | ✓ | ✓ | ✓ | | |
| [21] | Model perdictive control | homo | ✓ | ✓ | ✓ | | ✓ |
| [22] | Deep reinforcement learning | homo | | | ✓ | | |
| [23] | Deep reinforcement learning | homo | | ✓ | ✓ | ✓ | |
| [28] | Receding horizon optimization | hetero | ✓ | ✓ | | | |
| [30] | particle swarm optimization | hetero | ✓ | | ✓ | | |
| This work | Model predictive control | hetero | ✓ | ✓ | ✓ | ✓ | ✓ |

Abbreviation: maneuverability (maneu), communication (com), constraints (const)

## III. Description of cooperative search problem

In solving a cooperative search problem based on distributed MPC, an objective function is constructed based on the target existence probability, environmental uncertainty, and UAV collision avoidance factors to describe the revenue of search decisions quantitatively. The complete cycle optimization problem is transformed into a short cycle discrete optimization problem using the receding horizon, enabling the UAVs to have online distributed search decision capabilities.

### A. Construction of objective function

For distributed cooperative search strategies without the central node, the objective function of each UAV can be written as:

$$J_i(t) = w_1 J_P + w_2 J_E + w_3 J_C \\ t \in [0,T], \ i \in \{1, 2, \cdots, N_U\} \tag{1}$$

In (1), $T$ is the total duration of the cooperative search task, $N_U$ is the number of UAVs, $J_i$ is the decision revenue of UAV $i$, $J_P$ is the target search benefit, $J_E$ is the environmental search benefit, $J_C$ is the collision avoidance benefit, and $w$ is the weight coefficient. The expression of $J_P$ can be written as:

$$J_P(t) = \sum_{(x,y) \in A_i} (1 - \zeta(x,y,t)) p(x,y,t) \tag{2}$$

In (2), $A_i$ is the detection area of UAV $i$, $\zeta$ refers to the situation where UAVs independently determine the presence of targets in the grid, defined explicitly as:

$$\zeta(x,y,t) = \begin{cases} 1, & p(x,y,t) \geq \delta_P \\ 0, & other \end{cases} \tag{3}$$

In (3), $\zeta(x,y,t)=1$ represents the presence of a target at position $(x,y)$ and time $t$, $\zeta(x,y,t)=0$ represents the absence of a target, $\delta_P$ is the threshold at which the target exists, and $p(x,y,t)$ is the target existence probability at time position $(x,y)$ and time $t$. Assuming there are $N_T$ targets in the task area, based on prior information, the possible location of the targets is $[x_{ta}^n, y_{ta}^n]^T$, $n=1,2,\cdots,N_T$. Using Gaussian distribution to initialize the target existence probability, where the target position provided by prior information is at the peak of the Gaussian distribution, the initial target existence probability can be written as:

$$p(x,y,0) = \sum_{n=1}^{N} c_n \exp\left[-\frac{(x-x_{ta}^n)^2 + (y-y_{ta}^n)^2}{v_n^2}\right] \tag{4}$$

In (4), $v_n$ and $c_n$ represent the target existence probability's peak width and height, respectively. During the execution of cooperative search tasks, UAVs dynamically update the local target existence probability based on their sensor detection information. At the same time, Bayesian estimation is used to update the target existence probability $p(x,y,t)$ based on the sensor detection probability $P_D$ and false alarm probability $P_F$. The specific description is as follows [18]:

$$p(x,y,t+1) = \begin{cases} \dfrac{P_D p(x,y,t)}{P_F + (P_D - P_F) p(x,y,t)}, & \zeta = 1 \\ \dfrac{(1-P_D) p(x,y,t)}{1 - P_D p(x,y,t) - P_F (1 - p(x,y,t))}, & \zeta = 0 \end{cases} \tag{5}$$

In (1), the expression of $J_E$ can be written as:

$$J_E(t) = \sum_{(x,y) \in A_i} \chi(x,y,t) \tag{6}$$

In (6), $\chi(x,y,t)$ represents the environmental uncertainty at position $(x,y)$ and time $t$. For the undetected area set $\chi(x,y,t)=1$, the uncertainty of each detection coverage area changes to $\chi(x,y,t) = 0.5\chi(x,y,t)$.

In (1), the expression of $J_C$ can be written as:

$$J_C = 1 - \|\mathbf{F}_i(t)\|, \quad \begin{cases} \mathbf{F}_i(t) = \sum_{j \neq i} \mathbf{F}_{ij}(t) \\ \mathbf{F}_{ij}(t) = \begin{cases} k e^{-\mu D_{ij}} \mathbf{d}_{ij}, & D_{ij} \leq D_{\max} \\ 0, & other \end{cases} \end{cases} \tag{7}$$

In (7), $\mathbf{F}_i$ is the combined virtual repulsive force on UAV $i$, $\mathbf{F}_{ij}$ is the virtual repulsive force exerted by UAV $j$ on UAV $i$, $D_{\max}$ is the maximum effective distance between UAVs to generate repulsive force, $D_{ij}$ is the distance between UAV $j$ and UAV $i$, and $\mathbf{d}_{ij}$ is the unit vector from UAV $j$ to UAV $i$ direction.

**B. Distributed model predictive control**

In cooperative search, UAVs are treated as particles in space. Divide the rectangular task area into $L_x \times W_y$ discrete grids, with each grid labeled $(x_g, y_g)$ and $x_g \in \{1,2,\cdots,L_x\}$, $y_g \in \{1,2,\cdots W_y\}$, to discretize the motion of UAVs in a rasterized map. In order to make UAV's cooperative search decision more visionary, based on the current state, the distributed MPC method is used to plan search points for future time intervals. At the discrete moment $k \in \mathbb{N}^+$, the state and actions of the UAV $i$ can be expressed as:

$$\begin{cases} s_i(k) = (x,y)_k^i \\ u_i(k) \in \{-1,0,1\} \end{cases} \tag{8}$$

In (8), $u_i(k) = 0$ represents the UAV flying in a straight line, $u_i(k) = -1$ represents left turn, and $u_i(k) = 1$ represents right turn. According to $s_i(k)$ and $u_i(k)$, the state of UAV $i$ at time $k+1$ can be calculated as:

$$s_i(k+1) = f(s_i(k), u_i(k)) \tag{9}$$

In (9), function $f$ represents the mapping relationship between the current action and state to the next moment state. Assuming the number of prediction steps in the discrete-time domain is $m$, the predictive state sequence $s_i(k+1), s_i(k+2), \cdots, s_i(k+m)$ can be obtained sequentially through the current state $s_i(k)$ and the action sequence $u_i(k), u_i(k+1), \cdots, u_i(k+m-1)$. Taking moving one grid at a time interval as an example, when $m=3$, all predictive state sequences of the UAV are shown in Fig. 1.

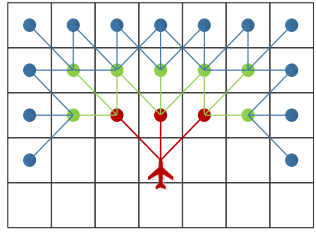

Fig. 1 Discrete predictive state sequences of UAV

The nodes with different colors in Fig. 1 represent the states at different times, and each red, green, and blue dot connected by a line represents a predictive state sequence. The direction options of each node represent the actions at the current time. The state and action sequence of a search trajectory predicted by the UAV $i$ at time $k$ can be expressed as:

$$\begin{aligned} \mathbf{S}_i(k) &= [s_i(k+1), s_i(k+2), \cdots, s_i(k+m)] \\ \mathbf{U}_i(k) &= [u_i(k), u_i(k+1), \cdots, u_i(k+m-1)] \end{aligned} \quad (10)$$

The UAVs adopt the MPC method for distributed autonomous decision-making, and each UAV shares its state, actions, and search results in the communication network. During each decision cycle, the UAV searches for the optimal solution among all predictive action sequences based on the information it can obtain. The cooperative search decision model for the UAV $i$ can be expressed as:

$$\begin{aligned} \mathbf{U}_i^*(k) &= \arg\max \hat{J}_i(\mathbf{U}_i(k), s_i(k), O_{-i}(k)) \\ s.t. &\begin{cases} s_i(k+q+1|k) = f_i(s_i(k+q|k), u_i(k+q|k)) \\ s_i(k+q+1|k) \in \Xi \\ u_i(k+q|k) \in \Theta \\ q \in \{0,1,2,\cdots,m-1\}, i \in \{1,2,\cdots,N_U\} \end{cases} \end{aligned} \quad (11)$$

In (11), $O_{-i}(k)$ represents the shared information of other UAVs except for UAV $i$ at time $k$. $\mathbf{U}_i^*(k)$ represents the optimal action sequence solved by the optimization problem. $\Xi$ and $\Theta$ represent the possible state and allowable action sets during decision-making, respectively. According to (11), UAV $i$ calculates the objective function with $\mathbf{U}_i(k)$ as the unit in the search decision. In the case where $s_i(k)$ is known, each $\mathbf{U}_i(k)$ corresponds to a unique $\mathbf{S}_i(k)$. According to (10), $\mathbf{S}_i(k)$ comprises several predictive states, and the UAV performs a search revenue calculation at each predictive state. Therefore, (11) can also be written as:

$$\hat{J}_i(\mathbf{U}_i(k), s_i(k), O_{-i}(k)) = \sum_{q=1}^{m} J_i(k+q) \quad (12)$$

In (12), $J_i(k+q)$ represents the search revenue of UAV $i$ at the predictive state $s_i(k+q)$, which is also the expression of (1) in the discrete-time domain. The expression of $J_i(k)$ can be written as:

$$J_i(k) = w_1 J_P + w_2 J_E + w_3 J_C$$
$$k \in \mathbb{N}^+; \ i = 1, 2, \cdots, N_U \quad (13)$$

Based on the state $s_i(k)$ and heading angle $\varphi_i(k)$, the detection area $A_i$ can be determined, and then the results of $J_P$, $J_E$ and $J_C$ in the objective function can be calculated. Define the predictive state sequence set and predictive action sequence set of UAV $i$ at time $k$ as $\bar{\mathbf{S}}_i(k)$ and $\bar{\mathbf{U}}_i(k)$, respectively, and meet the following requirements:

$$\begin{cases} \mathbf{S}_i(k) \in \bar{\mathbf{S}}_i(k) \\ \mathbf{U}_i(k) \in \bar{\mathbf{U}}_i(k) \end{cases} \quad (14)$$

According to (11) and (14), the UAV needs to select $\mathbf{U}_i^*(k)$ from the action sequence set $\bar{\mathbf{U}}_i(k)$ to maximize the value of the objective function $\hat{J}_i$. This paper uses $\mathbf{U}_i(k)$ as the genome, and the Genetic Algorithm is adopted to solve the optimization problem. After obtaining $\mathbf{U}_i^*(k)$, the first action $u_i^*(k)$ in the sequence is used as the expected control output of the UAV $i$, causing the UAV to move to the next search point. After the UAV reaches a new search point, repeat the above steps to generate a new control output.

## IV. Jump grid decision method

There are significant differences in the maneuverability of heterogeneous UAVs. In order to ensure that the search decision results meet the lateral acceleration and velocity constraints, a jump grid decision method is proposed in this paper. This method improves the definition of the possible state set $\bar{\mathbf{S}}_i(k)$ based on MPC, allowing UAVs to move between non-adjacent grids, maintaining the advantage of low computational cost in single variable discrete optimization problems, and satisfying the maneuverability constraints of UAVs.

### A. Determine of jump grid value

Define the possible state set of the UAV $i$ at time $k+1$ as $\bar{s}_i(k+1)$. Taking time $k$ as the current time, the corresponding relationship between the state of UAV $i$ and the grid position in Fig. 2 is shown in Table 2.

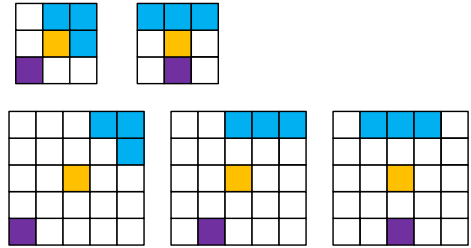

Fig.2 State sets with jump grid value equals 1 and 2

Table 2. UAV states and grid positions.

| State | Position | Meaning |
|---|---|---|
| $s_i(k-1)$ | purple grid | last state |
| $s_i(k)$ | yellow grid | current state |
| $\bar{s}_i(k+1)$ | blue grid | possible future states |

It can be seen from Fig .2 that when the jump grid value $j=1$, the grids of $s(k)$ and $\bar{s}(k+1)$ are adjacent, which is the same as traditional search decisions. When the jump grid value $j=2$, $s(k)$ and $\bar{s}(k+1)$ are not adjacent and need to move two grids from $s(k)$ to $\bar{s}(k+1)$. The predictive state sequence set for $m=3$ and $j=2$ is shown in Fig. 3.

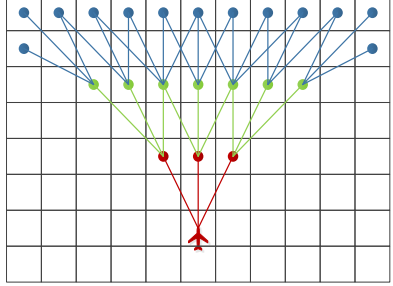

Fig. 3 Predictive state sequence set with jump grid value equals 2

Comparing Fig. 3 and Fig. 1, it can be found that under the same conditions of $m$ and $\bar{\mathbf{U}}_i(k)$, the spread angle of the state sequence set $\bar{\mathbf{S}}_i(k)$ with $j=2$ is smaller, equivalent to a smaller range in the heading angle of the UAV. In heterogeneous clusters, the flight speeds and lateral acceleration of different types of UAVs are different. To accurately select the parameters based on the UAV's maneuverability, the state sequences of $j=1,2,3$, $u(k)=-1$ (only maneuvering to the left) and the turning angles (changes of heading angle) of $j=3$ are shown in Fig. 4 and Fig. 5, respectively.

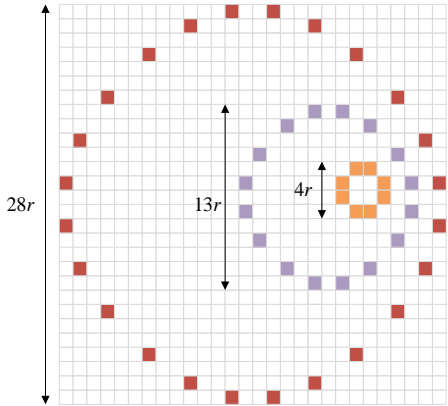 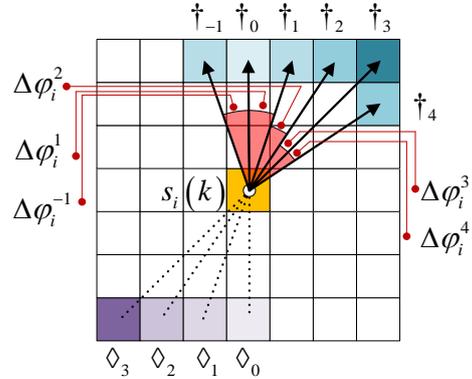

Fig. 4 Turning radius corresponding to different jump grid values

Fig. 5 Illustration of UAV turning angles

In Fig. 4, a closed curve resembling a circle can be formed by connecting grids with the same color. Since the UAV only maneuvers in one direction, the turning radius of the UAV corresponding to different jump grid values can be estimated as

$$R = \left(\frac{1}{2}(j+1)^2 + j(j-1)\right)r \quad (15)$$

In (15), $r$ is the side length of the square grid, and $R$ is the turning radius.

In Fig. 5, $\dagger_n = s_i^n(k+1)$ represents the predictive state at the next moment, $\lozenge_n = s_i^n(k-1)$ represents the state at the previous moment, and $n$ represents the grid label. $s_i^n(k-1)$ determines the heading angle ($\varphi_i^n$) of the UAV at the current moment. From Fig. 5, it can be seen that the turning angles satisfy $|\Delta\varphi_i^n| = |\Delta\varphi_i^{-n}|$ and $|\Delta\varphi_i^j| = |\Delta\varphi_i^{j+1}|$. Therefore, we only need to calculate $\Delta\varphi_i^n, n \in [1, j]$ to obtain the absolute values of all turning angles. The calculation method of $\Delta\varphi_i^n$ can be expressed as:

$$\Delta\varphi_i^n = \tan^{-1}\left(\frac{n}{j}\right) - \tan^{-1}\left(\frac{n-1}{j}\right) \tag{16}$$

Under the condition of setting the UAV to fly at a constant speed, first set the grid size $r$ based on the search accuracy requirements and then determine the minimum turning radius of the UAV as:

$$R_{\min} = V^2 / a_{\max} \tag{17}$$

In (16), $R_{\min}$ is the minimum turning radius of the UAV, $V$ is the flight speed, and $a_{\max}$ is the maximum lateral acceleration. Equations (15), (17) are introduced into $R > R_{\min}$ to meet the maneuverability constraints, and the range of $j$ is satisfied:

$$\frac{R_{\min}}{r} < \left(\frac{1}{2}(j+1)^2 + j(j-1)\right) \tag{18}$$

Under the condition of setting the UAV's search decision interval as constant, the lateral acceleration can be written as

$$a_n = V \cdot \frac{\Delta\varphi_n}{\Delta t} = \frac{r \cdot \sqrt{j^2 + n^2}}{\Delta t} \cdot \frac{\tan^{-1}\left(\frac{n}{j}\right) - \tan^{-1}\left(\frac{n-1}{j}\right)}{\Delta t} = k_a \cdot \frac{r}{(\Delta t)^2}$$
$$k_a = \sqrt{j^2 + n^2} \cdot \left(\tan^{-1}\left(\frac{n}{j}\right) - \tan^{-1}\left(\frac{n-1}{j}\right)\right) \tag{19}$$

In (19), $k_a$ is the acceleration coefficient. When $j = 1$, the turning angle of the UAV is only $45°$ (Fig. 2), and the acceleration coefficient is $\sqrt{2}\pi/4$. The turning angle and $k_a$ of the UAV when $j \in [2,10]$ are shown in Fig. 6 and Fig. 7, respectively.

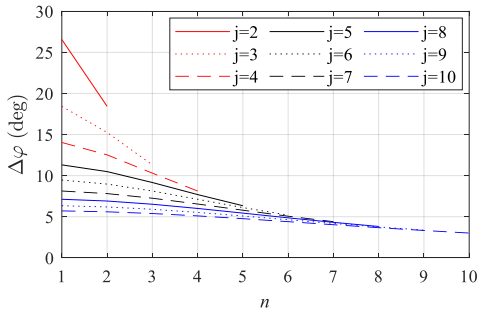 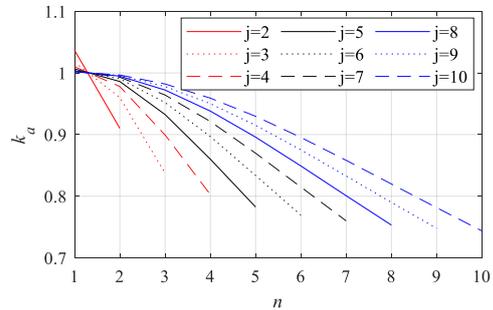

**Fig. 6 Turning angles**           **Fig. 7 Acceleration coefficients**

Figures 6 and 7 show that the maximum turning angle and maximum acceleration of the UAV occur at $n=1$. Since $k_a \leq \sqrt{2}\pi/4$, the grid size and decision interval can be preliminarily determined based on the UAV's maneuverability to satisfy

$$\frac{\sqrt{2}\pi}{4} \cdot \frac{r}{(\Delta t)^2} < a_{max} \tag{20}$$

Afterward, based on the velocity constraint of the UAV, determine the range of $j$ to satisfy

$$\frac{V_{min}\Delta t}{r} < j < \frac{\sqrt{2}V_{max}\Delta t}{2r} \tag{21}$$

In (21), $V_{max}$ and $V_{min}$ represent the UAV's maximum and minimum flight speeds, respectively.

**B. Update of the UAV search position**

In the jump grid decision method, different values of $j$ correspond to different possible state sets $\bar{s}_i(k+1)$. It is necessary to design the mapping relationship $f(s_i(k), u_i(k))$ in (9) to continuously update the state of UAVs in MPC. The $s_i(k)$ is numbered to enable $s_i(k) = (x, y)_k^i$ to perform mathematical operations with $u_i(k) \in \{-1, 0, 1\}$.

Firstly, take the grid where the UAV $i$ is located as the center and number all the grids on the layer where the jump grid is located from $-4j$ to $+4j$. Define the grid label as $n_i$, where the grid with $n_i = 0$ is located at the heading angle of 0, and the grid with $n_i = \pm 4j$ represents the same grid at the heading angle of $\pm \pi$. Taking $j = 2$ as an example, the numbering method for jump grids is shown in Fig 8.

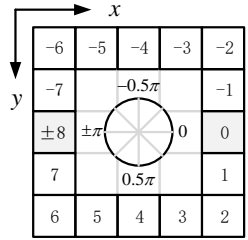

Fig. 8 Numbering method for jump grids

Secondly, determine the jump grid number based on the UAV's heading angle. The heading angle interval corresponding to the jump grid number is shown in (22).

$$\varphi_i(k) \in \left[\frac{45(2n_i(k)-1)}{2j}, \frac{45(2n_i(k)+1)}{2j}\right] \tag{22}$$

In (22), $\varphi_i(k)$ represents the heading angle of the UAV $i$ at time $k$ in the discrete-time domain, and $n_i(k)$ represents the grid number. According to $n_i(k)$ and $u_i(k)$, the grid number for the next moment can be calculated as:

$$n_i(k+1) = n_i(k) + u_i(k) \tag{23}$$

After obtaining grid number $n_i(k+1)$ at the next moment, it is necessary to calculate the predictive state of UAV $s_i(k+1)$ in reverse. The increments of grid coordinates in the $x$ and $y$ direction corresponding to different grid numbers are shown in Fig. 9 and Fig. 10, respectively.

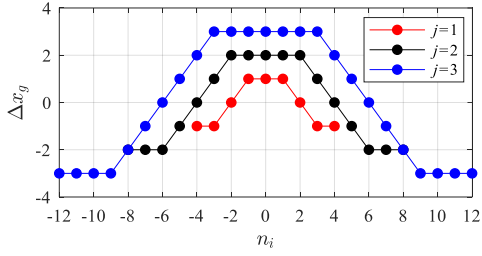
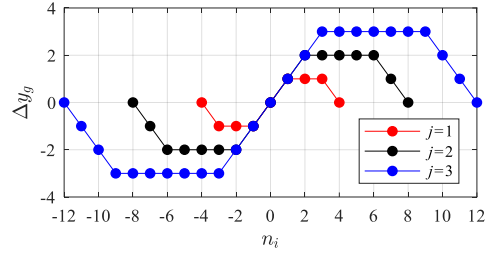

**Fig. 9 The increments of grid coordinate in the x-direction**

**Fig. 10 The increments of grid coordinate in the y-direction**

In Figs. 9 and 10, the horizontal axis represents the grid number, reflecting the UAV's motion direction, while the vertical axis represents the increment of grid coordinates in the $x$ and $y$ directions at the next moment compared to the current moment. According to Figs. 9 and 10, the relationship between $n_i$ and ($\Delta x_g$, $\Delta y_g$) is summarized as shown in (24).

$$\Delta x_g = \begin{cases} -j, & n_i \in [-4j,-3j] \\ 2j+n_i, & n_i \in [-3j,-j] \\ j, & n_i \in [-j,j] \\ 2j-n_i, & n_i \in [j,3j] \\ -j, & n_i \in [3j,4j] \end{cases}, \quad \Delta y_g = \begin{cases} -n_i-4j, & n_i \in [-4j,-3j] \\ -j, & n_i \in [-3j,-j] \\ n_i, & n_i \in [-j,j] \\ j, & n_i \in [j,3j] \\ -n_i+4j, & n_i \in [3j,4j] \end{cases} \quad (24)$$

The increment of grid coordinates from time $k$ to $k+1$ can be obtained by bringing $n_i(k+1)$ into (24), and then the grid coordinates at time $k+1$ can be calculated as:

$$(x_g, y_g)_{k+1}^i = (x_g, y_g)_k^i + (\Delta x_g, \Delta y_g)_k^i \quad (25)$$

According to the grid coordinates, the actual position coordinates of the UAV can be calculated as follows:

$$\begin{cases} x = r(x_g - 0.5) \\ y = r(y_g - 0.5) \end{cases} \quad (26)$$

The predictive state $s_i(k+1) = (x,y)_{k+1}^i$ of UAV $i$ at time $k+1$ can be obtained.

## V. Parameter dynamic selection method

According to Sections III and IV, the adjustable parameters in the cooperative search include the jump grid value ($j$), the prediction sequence length ($m$), and objective function weights ($w_i$). A parameter dynamic selection method is designed to make search decisions adaptive to meet the needs of tasks.

### A. Search parameters online selection based on expert systems

Due to the different orders of magnitudes of $J_P$, $J_E$, and $J_C$ in the objective function, normalizing the weight values makes the comparison of search benefits less intuitive. Therefore, the parameters designed for online selection in the expert system are $j$, $m$, $k_{w1}$, $k_{w2}$, $k_{w3}$, where $k_{w1}$, $k_{w2}$, $k_{w3}$ are weight correction coefficients, and the usage method is as follows.

$$w_i = k_{wi} \cdot w_i \quad i = 1, 2, 3 \tag{27}$$

The closest distance from the UAV $i$ to targets at time $k$ is defined as:

$$b(k) = \min\left(\left|(x, y)_k^i - (x_t, y_t)_k^l\right|\right) \tag{28}$$

In (28), $(x, y)_k^i$ represents the position of UAV $i$ at time $k$, $(x_t, y_t)_k^l$ represents the position of target $l \in [1, N_T]$ at time $k$, and $N_T$ represents the number of targets known based on prior information. When the UAV has not discovered the target, the target information comes from prior information, which is an inference and prediction of the target state.

Design expert systems to select search parameters for UAVs and define the input of Expert System 1 as:

$$E_1(k) = b_i(k)/b^* \tag{29}$$

In (29), $b^*$ represents the target warning distance, which is manually set. The input for Expert System 2 is defined as:

$$E_2(k) = N_T^*(k)/N_T \tag{30}$$

In (30), $N_T^*(k)$ is the number of targets discovered by UAVs at time $k$. The design of the expert systems is shown in Table 3.

**Table 3 Expert systems**

| Expert system 1 | | | Expert system 2 | | | |
|---|---|---|---|---|---|---|
| $E_1(k)$ | $j$ | $k_{w3}$ | $E_2(k)$ | $k_{w1}$ | $k_{w2}$ | $m$ |
| $[0,1)$ | 2 | 2 | $[0, 0.8)$ | 1 | 1 | 8 |
| $[1,2)$ | 4 | 1 | $[0.8, 1)$ | 0.8 | 1.2 | 10 |
| $[2,\infty)$ | 6 | 0.8 | $[1,\infty)$ | 0.4 | 1.6 | 12 |

In Table 3, the parameter settings of the expert systems are for reference only, and these parameter values can be appropriately modified in different task scenarios. The focus of designing expert systems is to ensure that the parameter adjustment rules meet the requirements of the current task state. Under the same search decision interval, the function of Expert System 1 can be described as follows: When the UAV is far away from the target, select a larger jump grid value and quickly fly over areas with low target existence probability. When the UAV is close to the target, select a smaller jump grid value, and reduce flight speed and turning radius to search the area carefully. Besides, a more considerable collision avoidance weight is selected to prevent UAVs from gathering in areas with a high target existence probability. The function of Expert System 2 can be described as follows: When the proportion of target discovery is low, prioritize searching for areas with a high target existence probability to discover targets as soon as possible; When the proportion of target discovery is large, prioritize searching for areas with high environmental uncertainty to avoid target omission due to inaccurate prior information. Besides the prediction sequence length is additionally increased to make the UAV decision-making more visionary.

**B. Maneuverability dynamic adjustment strategy**

The threat areas in cooperative search tasks include areas outside the search boundary and denied areas (obstacles) within the boundary. When approaching the threat area, UAVs must avoid threats to complete search tasks safely. When conducting threat avoidance, the UAV is expected to maintain its original flight state and gradually reduce its flight speed and turning radius. Therefore, in the process of threat avoidance, it is not suitable to use expert systems to adjust jump grid value directly. It is necessary to design a method for continuously adjusting the flight states of UAVs. Under the same search decision interval, the dynamic adjustment strategy for UAV's maneuverability is as follows.

(1) Determine the predictive action sequence set $\bar{\mathbf{U}}_i(k)$ based on the output of $j$ and $m$ from expert systems.

(2) Using genetic algorithms to solve the optimization problem shown in (11). $N_g$ genes ($\mathbf{U}_i(k)$) are randomly generated as the genome, and $\mathbf{S}_i(k)$ is calculated through $\mathbf{U}_i(k)$. It can be inferred that $\mathbf{S}_i(k)$ contains the predictive position of the UAV, according to (8) and (10).

(3) Check the predictive position of the UAV. When the predictive position of the UAV enters the threat area, $s_i(k+q+1|k) \notin \Xi$ does not meet the constraint conditions in (11) and sets $\hat{J}_i(\mathbf{U}_i(k), s_i(k), s_{-i}(k)) < 0$.

(4) After solving with the genetic algorithm, if there is no $\mathbf{U}_i(k)$ that makes $\hat{J}_i > 0$, it indicates that the flight speed and turning radius corresponding to the current jump grid value cannot ensure that the UAV avoids threat areas. Make $j = j-1$, decrease the jump grid value, and repeat the above steps until $\mathbf{U}_i(k)$ exists to make the $s_i(k+q+1|k) \in \Xi$ condition valid.

The dynamic adjustment process of UAV's maneuverability is shown in Fig. 11, where $c$ represents the number of genetic algorithm iterations, $n$ represents the maximum number of iterations, $\varepsilon$ represents the termination error, and $\hat{J}_i^*$ represents the maximum value of the objective function calculated by the genetic algorithm.

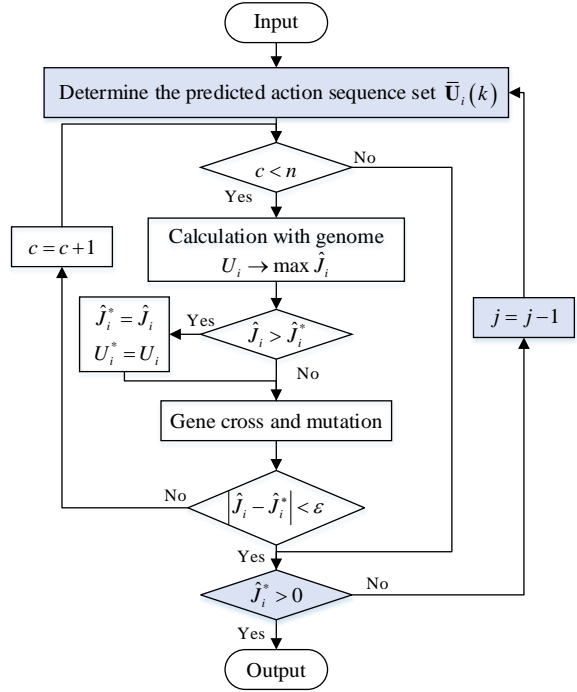

Fig. 11 Flow chart of maneuverability dynamic adjustment process

## VI. Search information transmission method

In distributed cooperative search, UAVs obtain search information from others through data links to avoid repeated searches of task areas. In practical applications, external interference, communication bandwidth and range limitations may cause transmission interruption. Although the UAVs can continue to work in this situation, the lack of communication between individuals can decrease the cluster's search efficiency. A search information transmission method under communication constraints is designed to address the above problem, which achieves the centralized transmission of historical search information under small communication bandwidth.

The communication network of heterogeneous UAVs in the cooperative search task is shown in Fig. 12. In the figure, the dashed line represents the datalink between the rotor UAVs, and the solid line represents the datalink between the fixed-wing UAV and rotor UAVs, which has a more extended communication range. The Fixed-wing UAV can serve as a relay node to transmit search information between rotor UAVs. This approach can give full play to the advantages of the fixed-wing UAV in fast flight speed and wide communication range, effectively promoting the information flow among the rotor UAVs and improving the search performance of the cluster.

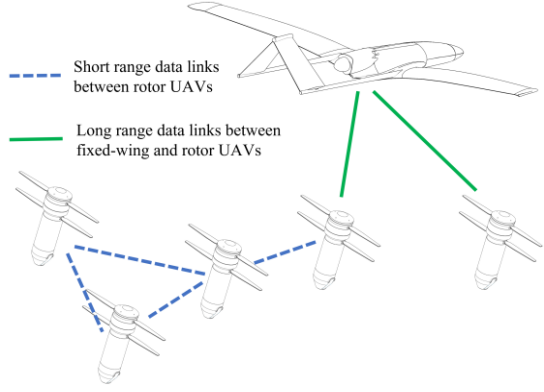

**Fig. 12 The communication network of heterogeneous UAVs**

**A. Design of information transmission content**

It can be seen from (2) (5) (6), that the UAV updates its target existence probability and environmental uncertainty based on the detection results, and calculates $J_p$ and $J_E$ based on $p(x,y,t)$ and $\chi(x,y,t)$. In cooperative search tasks, UAV $i$ needs to synchronously update its own $p_i(x,y,t)$ and $\chi_i(x,y,t)$ based on the shared information $O_{-i}(k)$ of other UAVs. Since $p_i(x,y,t)$ and $\chi_i(x,y,t)$ are functions of position $(x,y)$ and time $t$, $p_i(x,y,t)$ and $\chi_i(x,y,t)$ need to be transmitted in matrix form. The number of rows and columns in the matrix is consistent with the grid map size, and the matrix's element values change over time. If the matrix information of $p_i(x,y,t)$ and $\chi_i(x,y,t)$ is directly included in $O_{-i}(k)$, it will cause excessive data transmission.

The $O_{-i}(k)$ in (11) can be composed of many forms. Under the condition that UAV $i$ knows other UAVs' field of view, $O_{-i}(k)$ only needs to include the search position, heading angle, and target detection results of other UAVs so that UAV $i$ can reproduce the search results locally and synchronously update $p_i(x,y,t)$ and $\chi_i(x,y,t)$. This form of $O_{-i}(k)$ can greatly reduce the size of the message package. In addition, when communication bandwidth allows, historical search information can be transmitted, enabling UAVs to quickly obtain other UAVs' search information after re-establishing links and updating $p_i(x,y,t)$ and $\chi_i(x,y,t)$ locally. The design of message package and local information storage is shown in Fig. 13.

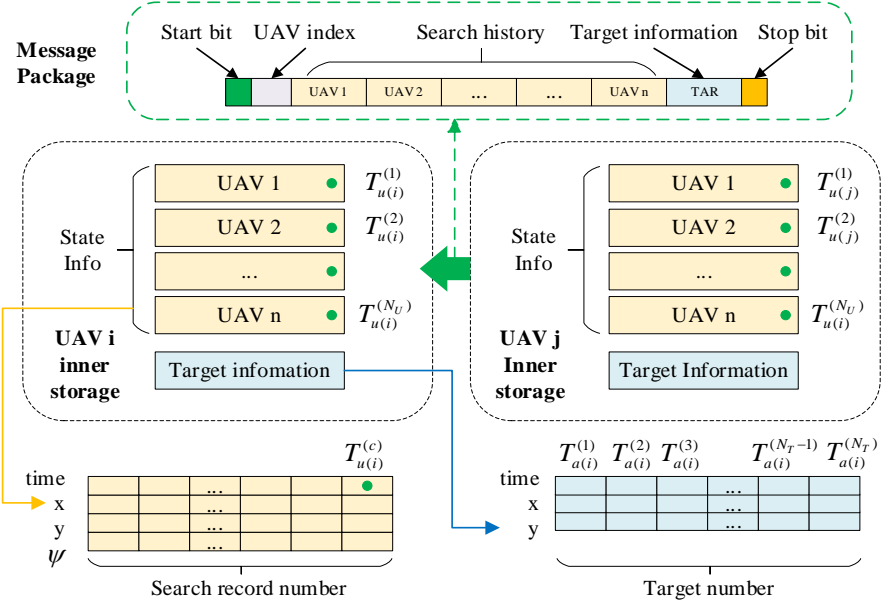

**Fig. 13 Message package and local information storage of the UAV**

For each UAV, the number of locally stored state information matrices (yellow) is $N_U$, and the number of target information matrices (blue) is 1. The state information matrix stores positions, heading angles, and corresponding timestamps of the UAV, while the target information matrix stores positions and corresponding timestamps of targets. During the communication process, the state information and target information are sequentially encoded and stored in a message package to achieve the sending and receiving of search information.

Given an example, there are 5 UAVs in the cluster, each of which stores ten target information and centrally transmits search information from the past 100 moments. According to Fig. 13, there are 400 elements in the state information matrix and 30 elements in the target information matrix. Each UAV stores five state information matrices and one target information matrix. All elements are stored in float (4bit) format. The message size for each communication is (400×5+30) 4=8.12kb.

**B. Update of local search information**

Due to the different communication ranges of UAVs in the heterogeneous cluster, UAV $i$ will receive messages sent by UAV $j$ when UAV $i$ enters the communication range of UAV $j$. The message contains the state information of UAVs and target information stored by UAV $j$. After receiving the message, UAV $i$ first reads the latest timestamp $T_{u(j)}^{(c)}$, $c \in [1, N_U] \cap \mathbb{N}^+$ of each UAV's state information in the message and compares the timestamps with the locally stored timestamps. Suppose $T_{u(j)}^{(c)} > T_{u(i)}^{(c)}$, the information of UAV $c$ in the message is used to replace the locally stored information of UAV $c$. Similarly, the timestamp $T_{a(j)}^{(d)} > T_{a(i)}^{(d)}$, $d \in [1, N_T] \cap \mathbb{N}^+$ determines whether to update the target information stored in the UAV $i$. The local information update method is shown in Fig. 14.

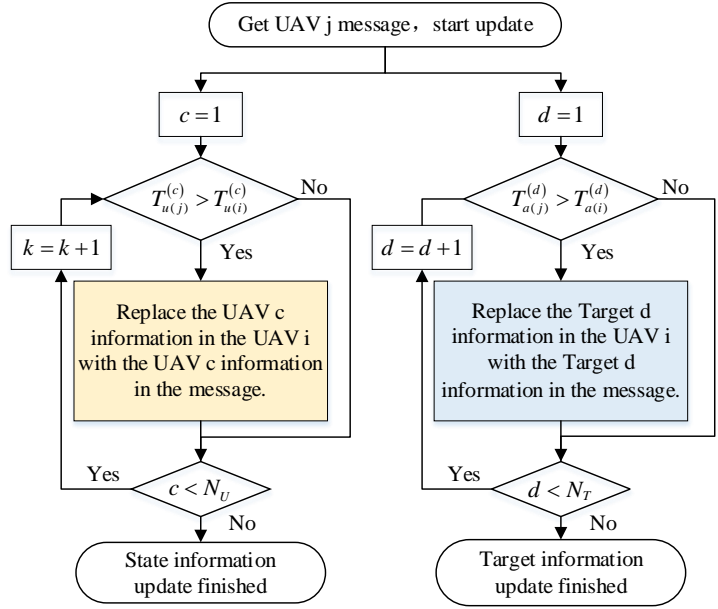

**Fig. 14 Local search information update method**

After the information transmission is completed, $p_i(x,y,t)$ and $\chi_i(x,y,t)$ are updated through (3) (5), based on the other UAVs' field of view, which are pre-stored locally. Finally, the UAV $i$ calculates the objective function value of the current state using (2) (6) (7) (13) based on the $p_i(x,y,t)$, $\chi_i(x,y,t)$, and other UAVs positions.

## VII. Simulations

In simulations, set the task area to 1600m × 800m, which includes five targets and six denied areas. The task area is designed to simulate a block scenario. The denied areas, with a diameter of 80m~200m, simulate the buildings or threat range of soldiers. The parameters of fixed-wing and rotor UAVs are shown in Table 4. Among them, the "com-distance" represents the communication distance under interference, and the "perc-distance" represents the perception distance for denied areas.

**Table. 4 parameters of heterogeneous cluster**

| Parameter | Rotors | Fixed-wing |
| --- | --- | --- |
| fligt height | 40m | 200m |
| flight speed | 0-35m/s | 20-70m/s |
| com-distance | 160m | 300m |
| perc-distance | 300m | 600m |
| field of view | 40m × 40m | - |

The parameter settings in the cooperative search approach are shown in Table 5.

**Table. 5 Parameters of cooperative search approach**

| Parameter | Value | Originate |
| --- | --- | --- |
| $c_n$ | 0.3 | Eq.(4) |
| $v_n$ | 50m | Eq.(4) |
| $p_D$ | 0.8 | Eq.(5) |
| $P_F$ | $1\times10^{-4}$ | Eq.(5) |
| $D_{max}$ | 200m | Eq.(7) |
| $k$ | 10 | Eq.(7) |
| $\mu$ | $6\times10^{-3}$ | Eq.(7) |
| $r$ | 4m | Eq.(15) |
| $b^*$ | 160m | Eq.(29) |

The genetic algorithm (GA) parameters for rotor UAVs and the fixed-wing UAV in the simulation are shown in Table 6. The gene length is equal to the number of prediction steps and is continuously corrected by the parameter dynamic selection method during the search process. Therefore, the gene length of rotor UAVs ranges from 6 to 10, while that of the fixed-wing UAV ranges from 10 to 15. Usually, the longer the genes, the greater the demand for population size, and they are exponentially related. Fortunately, in the simulation, the fixed-wing UAV only serves as a communication relay node and does not need to calculate search revenue. Therefore, for the fixed-wing UAV, genetic algorithms do not need to calculate the optimal solution; they only need to come up with feasible solutions to ensure flight safety. So, the population of the fixed-wing UAV is set to 300 to save computational resources. Its mutation rate and crossover rate can be chosen to be relatively large without considering the convergence of GA's optimal solution. In addition, an elite selection strategy is used in the GA solutions for both types of UAVs to guarantee that the solution quality will not degrade from one generation to the next.

**Table. 6 Parameters of Genetic Algorithm**

| Parameter | Rotors | Fixed-wing |
|---|---|---|
| population size | 100 | 300 |
| max generation | 50 | 50 |
| mutation rate | 0.5 | 0.9 |
| crossover rate | 0.5 | 0.9 |
| gene length | 6-10 | 10-15 |

Before the task starts, UAVs can obtain prior information, but the specific location of targets and denied areas are unknown. The location of denied areas, prior information of targets, and communication range of UAVs are shown in Fig. 15. The contour lines in the figure represent the distribution of the target existence probability.

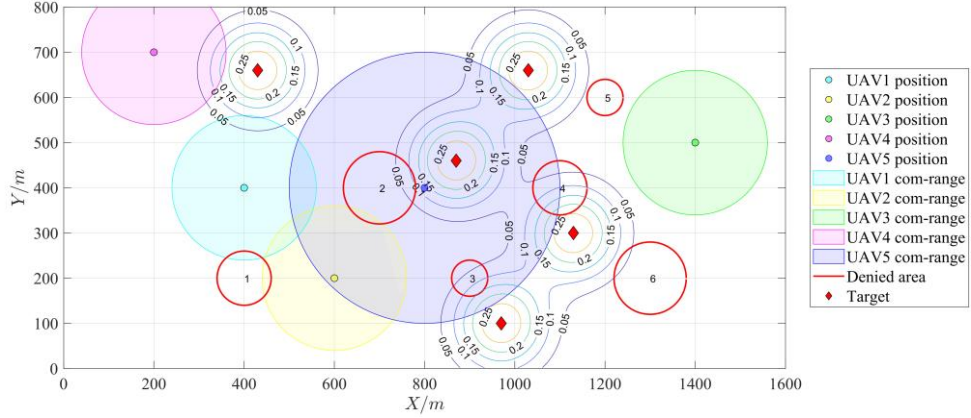

**Fig.15 Diagram of cooperative search task environment**

The simulation is conducted on a computer using an i7 CPU, and all UAVs' programs run in single MATLAB application. Thus, they jointly occupy the computing resources of one core in the processor. However, in the simulation, the decision-making of each UAV is decentralized, and the UAV obtains the state information and search information of other UAVs through shared memory to carry out distributed cooperative search tasks.

**A. Homogeneous UAVs cooperative search**

In the simulation of homogeneous UAVs, four rotor UAVs are set to search the task area, with a time interval of 1 second for each sensor detection. To verify the robustness of distributed cooperative search method, UAV 4 is set to suddenly exit the task after 100 search intervals (100 seconds after the task started), and the remaining UAVs continue to complete the task.

Under the condition of no communication constraints, the search result of the homogeneous cluster after 300 seconds of task start is shown in Fig. 16. The coverage areas of different colors in the figure are the search areas for different UAVs. The simulation result shows that UAVs discovered all targets within 300 seconds, reflecting the advantage of probability first in distributed cooperative search methods. Since the cluster is always in a fully connected state, each UAV can obtain search information from other UAVs timely, resulting in a slight repetition of search areas and high search efficiency.

From Fig. 16, it can also be seen that the search distance interval of UAVs in different regions is different. The change of the jump grid value causes the difference in search distance interval. In areas with a high target existence probability, the UAVs have a small jump grid value and low flight speed, so they carefully search targets; In areas with low target existence probability, UAVs have a considerable jump grid value and quickly fly over low probability areas. When the UAV approaches threat areas, the appropriate turning radius is obtained by gradually reducing the jump grid value, facilitating threat avoidance. The above phenomenon reflects the effectiveness of the parameter dynamic selection method.

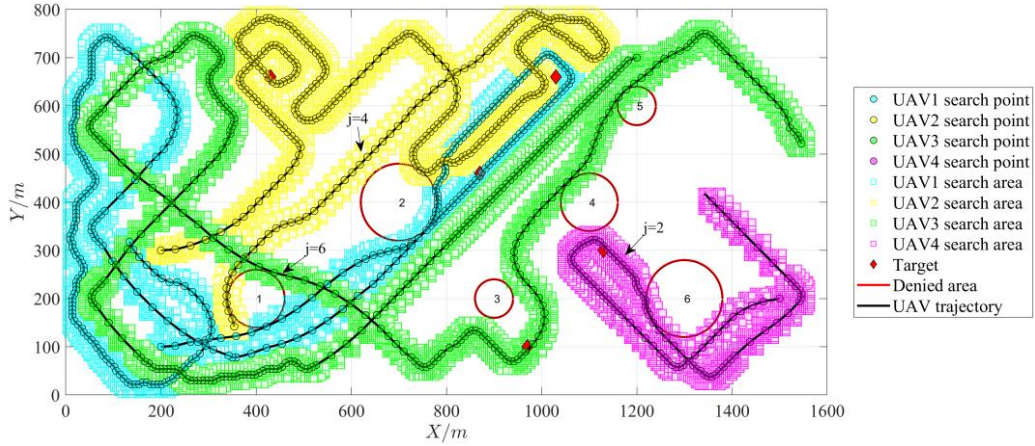

**Fig. 16 Cooperative search result of the homogeneous UAVs without communication constraints. More details can be found in the attached video.**

The search state is composed of environmental uncertainty and target existence probability. The search states of individuals and global are shown in Fig. 17. Since the cluster is fully connected, the search state of each UAV remains consistent with the global situation, and they gradually decrease as the search time increases. In 100 seconds, UAV 4 exited the search task, and its search state changed to 0. From Fig. 17 (a), the slope of the environmental uncertainty is close to a constant, indicating that the repeated search area of the cluster is small. From Fig. 17 (b), the target existence probability rapidly decreases and then tends to flatten out, reflecting the character of the cluster prioritizing the search for high-probability areas.

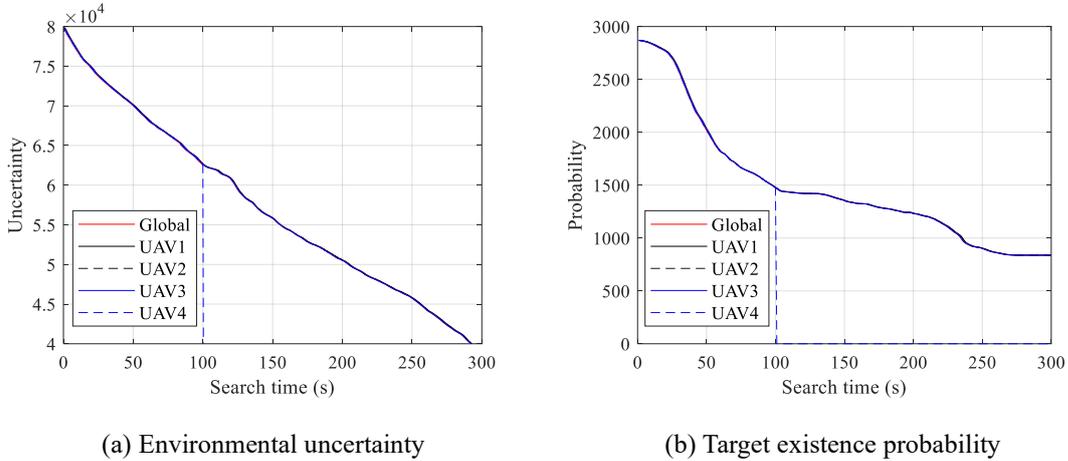

(a) Environmental uncertainty          (b) Target existence probability

**Fig. 17 Cooperative search state of homogeneous UAVs without communication constraints**

The expert system output and search revenue of each UAV are shown in Figs. 18 and 19, respectively. In Fig. 19, the solid line represents the weighted calculation results of $\mathbf{U}_i^*(k)$ in the objective function, while the dashed line represents the maximum and minimum values of the weighted results calculated by $\mathbf{U}_i(k) \in \bar{\mathbf{U}}_i(k)$. The figure shows that the solid line coincides with the upper dashed line representing the maximum value, indicating the correct optimization decision result. Based on Figs. 18 and 19, it indicates that in the early stage of the cooperative search, UAVs focus on the benefit of target existence probability, with a higher value of $w_1 J_P$; In the later stage of the search, UAVs focus on the benefits of environmental uncertainty, with $k_{w1}$ decreasing and $k_{w2}$ increasing, making $w_2 J_E$ becomes the main factor affecting search decisions.

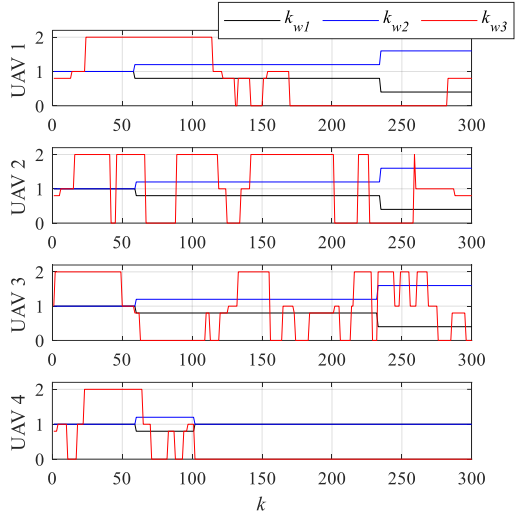 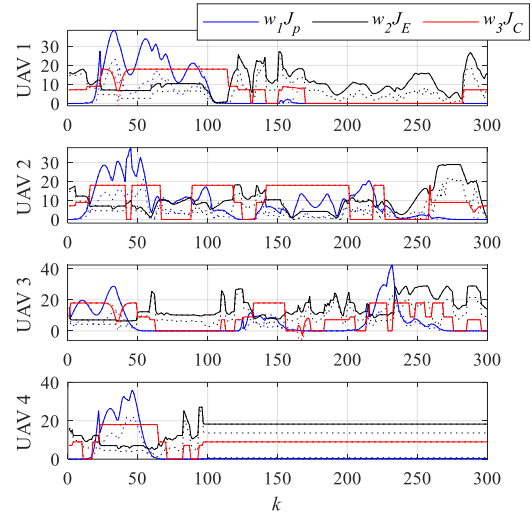

**Fig. 18 Expert system outputs**          **Fig. 19 Benefits of UAV cooperative search**

The cooperative search results of UAVs under communication constraints are shown in Fig. 20. From the figure, the cluster has experienced a lot of repeated searches in the task area and failed to discover all targets within 300s. This phenomenon is caused by the inability of UAVs to obtain search information from others timely, which reduces the search efficiency of the cluster.

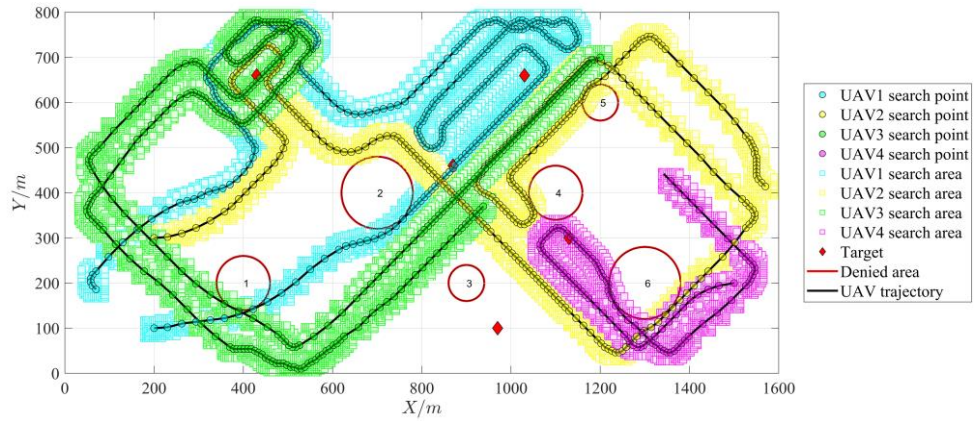

**Fig. 20 Cooperative search result of homogeneous UAVs with communication constraints. More details can be found in the attached video.**

Under communication constraints, the search states of individual and global UAVs are shown in Fig. 21. From the figures, there are step changes in the search state of each UAV, and there is a significant difference between the individual and global states. The step changes in the search state are caused by the UAV obtaining historical search information from others and replicating the search results locally. Therefore, it can also be seen from Fig. 21 that under communication constraints, the UAVs in the cluster only communicate at 29 seconds and 87 seconds.

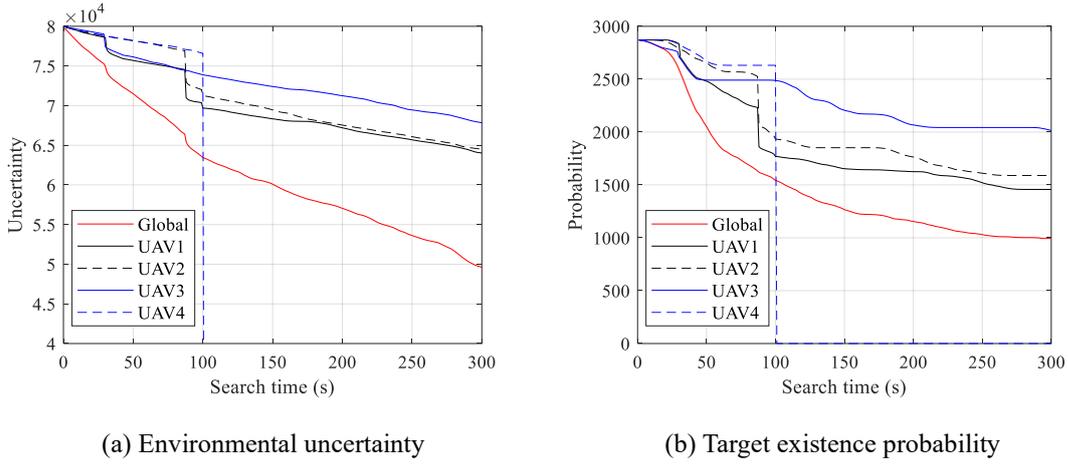

(a) Environmental uncertainty  (b) Target existence probability

Fig. 21 Cooperative search state of homogeneous UAVs with communication constraints

Fig. 17 and Fig. 21 indicate that under the same search time, the search efficiency of the cluster with communication constraints is lower than those without communication constraints. This phenomenon is due to the decrease in communication frequency, resulting in a decrease in the rationality of UAVs' search decisions. A heterogeneous UAVs cooperative search strategy is proposed to solve the above problems. The strategy introduces a fixed-wing UAV as the communication relay node to increase the communication frequency of rotor UAVs and improve the cooperative search efficiency of the cluster.

**B. Heterogeneous UAVs cooperative search**

Under the condition of homogeneous UAVs cooperative search task, a fix-wing UAV is added as a communication relay node. Usually, rotor UAVs are deployed in the task area by the fixed-wing UAV, so the heterogeneous UAVs search strategy does not create additional hardware requirements for cooperative tasks. At the beginning of the task, the jump grid value for the fixed-wing UAV is set to 12 and adjusted online during the threat avoidance process. The three-dimensional view of heterogeneous UAVs cooperative search result is shown in Fig.22. The figure shows that the fixed-wing UAV hovers over the task area and can automatically reduce its flight speed and turning radius when approaching the boundaries. The rotor UAVs' search area overlap is relatively small, and all targets are discovered within 300s.

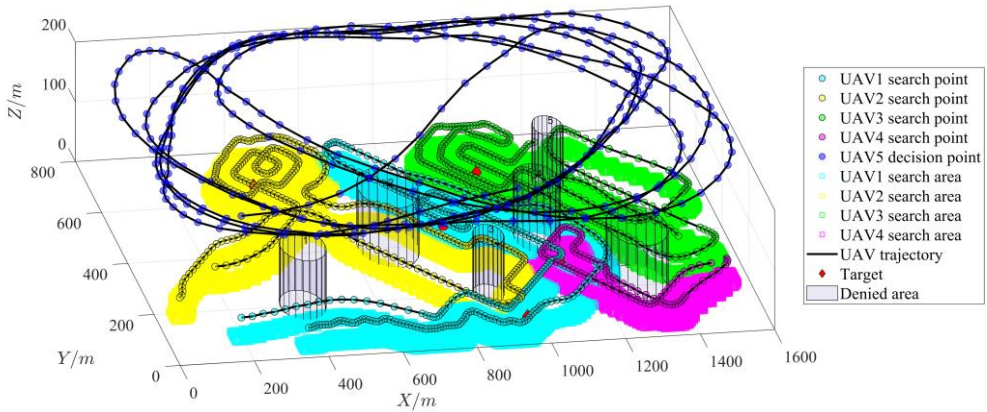

**Fig. 22 Cooperative search result of heterogeneous UAVs with communication constraints. More details can be found in the attached video.**

Under communication constraints, the heterogeneous cluster's individual and global search states are shown in Fig. 23. Figure shows that UAVs' search states have undergone multiple step changes, indicating that there have been multiple communications between the UAVs. The addition of a fixed-wing UAV has greatly improved the information transmission within the cluster. According to Fig. 21 and Fig. 23, the environmental uncertainty and target existence probability of the heterogeneous cluster decrease faster than the homogeneous cluster, indicating that the search efficiency of the heterogeneous cluster is significantly improved compared to the homogeneous cluster under communication constraints.

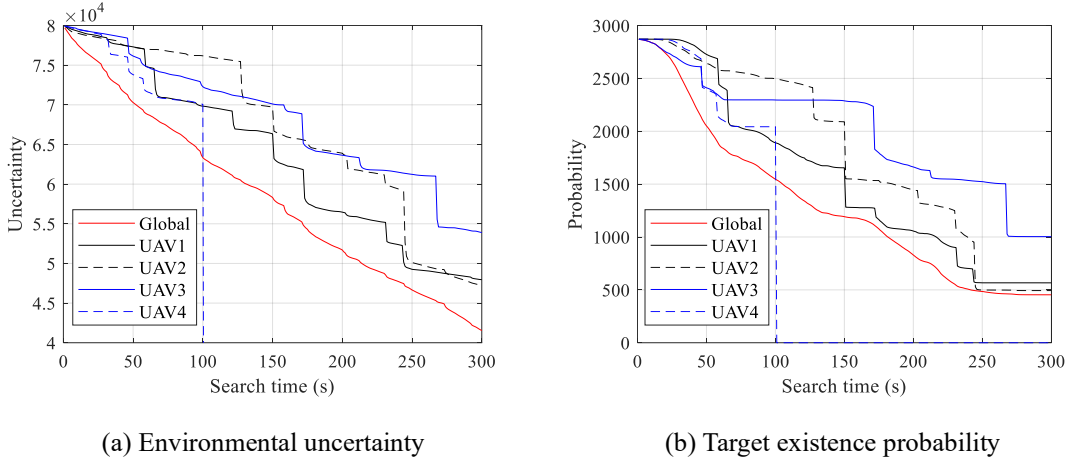

(a) Environmental uncertainty  (b) Target existence probability

**Fig. 23 Cooperative search state of heterogeneous UAVs with communication constraints**

### C. Comparison with coverage path planning method

Compare the distributed MPC-based search method proposed in this work with the coverage path planning (CPP) method [4][9] to verify the advantages of the MPC method in search efficiency. The parameters in Tables 4-6 in the simulation remain unchanged, and five polygon denied areas (obstacles) and five targets are set in the task area. To ensure fairness in comparison, set the UAV's field of view to a $l \times l$ rectangle in the MPC simulation and set the field of view width (flight path spacing) to $\sqrt{2}l$ in the CPP simulation. Stop the simulation when all targets are discovered. The simulation results are shown in Fig. 24.

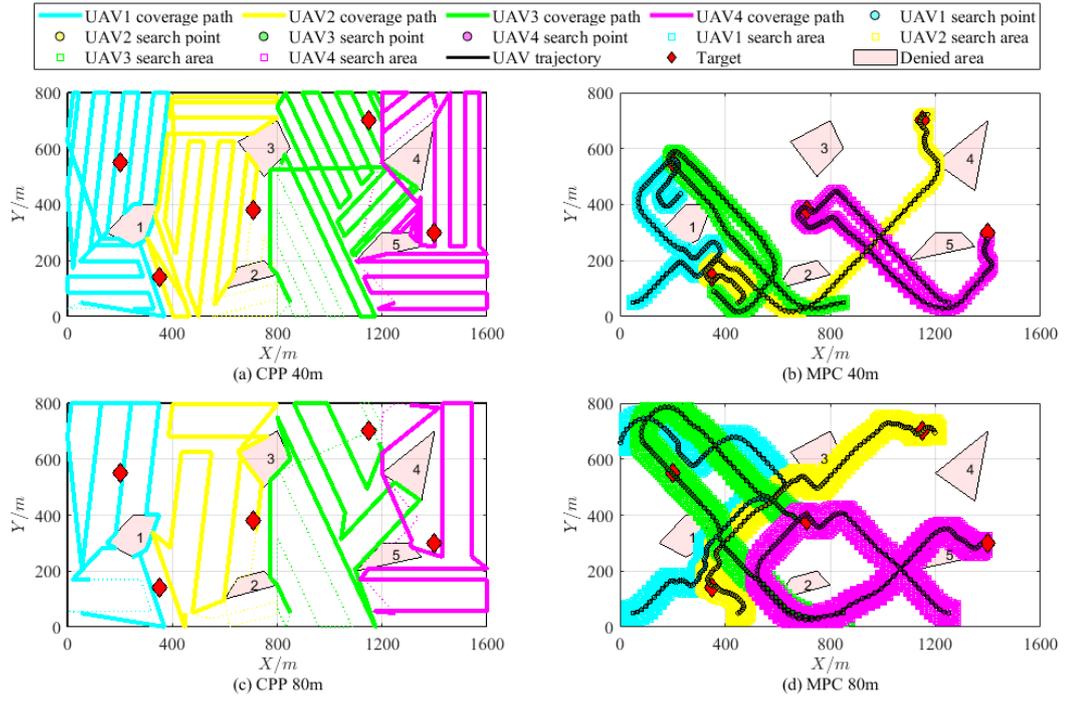

Fig. 24 Trajectory comparison of different search methods

Fig. 24 (a) and (b) show the search paths for CPP and MPC when $l = 40\text{m}$, respectively, and Fig. 24 (c) and (d) show the search paths for CPP and MPC when $l = 80\text{m}$, respectively. The figure shows that the total trajectory length of the MPC method is significantly shorter than that of the CPP method. The total trajectory length of the cluster corresponding to Fig. 24 is shown in Case 1 of Table 7. According to [4][9], the coverage path can be determined when the task area, denied area and path spacing are determined. Therefore, adjusting the target position can reduce the total trajectory length of the CPP method. Set up two target distribution schemes, and their simulation results are shown in Case 2 and Case 3 in Table 7, respectively.

Table 7 Comparison of cluster's total trajectory length

| Targets | $l$ (m) | MPC (m) | CPP (m) |
|---|---|---|---|
| Case 1 | 40 | 6729 | 26872 |
|  | 80 | 7011 | 15012 |
| Case 2 | 40 | 6580 | 17148 |
|  | 80 | 7527 | 13573 |
| Case 3 | 40 | 7982 | 6964 |
|  | 80 | 7835 | 9528 |

Table 7 shows that the distribution of targets has a relatively small impact on the total trajectory length of the MPC method. Moreover, in most cases, the total trajectory length of MPC is smaller than that of CPP. Therefore, the MPC method has significant advantages in quickly discovering targets. Fig. 24 shows that the trajectory repeatability planned by the CPP method is less than that of MPC, indicating that the CPP method has a significant advantage in the coverage efficiency of the task area.

**D. Evaluation of cooperative search strategy**

In solving optimization problems using genetic algorithms, the selection of $\mathbf{U}_i(k)$ is partially random, and the solution is suboptimal. Therefore, under the same initial conditions, there are differences in the results of each simulation. In order to evaluate the performance of different search strategies, the variation of the average search state is obtained through multiple simulations. Under the same initial conditions, three strategies are adopted for 900 seconds simulations. They are homogeneous UAVs without communication constraints (Strategy 1), homogeneous UAVs with communication constraints (Strategy 2), and heterogeneous UAVs with communication constraints (Strategy 3). Each strategy is simulated ten times, and their average environmental uncertainty and average target existence probability are shown in Fig. 25 and Fig. 26.

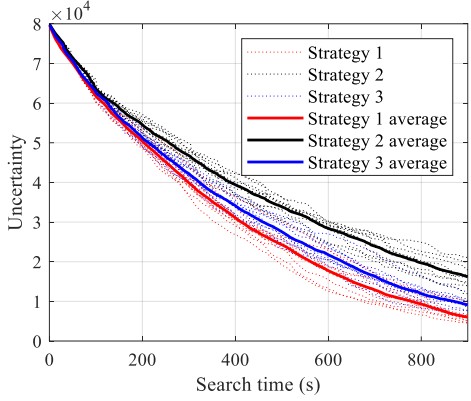
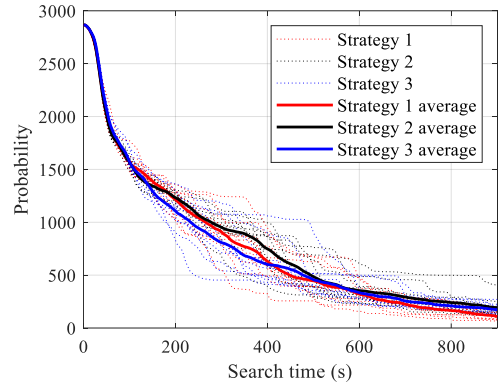

**Fig. 25 Environmental uncertainty (different strategies)**    **Fig. 26 Target existence probability (different strategies)**

Fig. 25 shows that the average environmental uncertainty of Strategy 1 decreases the fastest because UAVs can continuously share search information without communication constraints, avoiding repeat searches. Under communication constraints, the average environmental uncertainty decrease rate of Strategy 3 is close to Strategy 1 and significantly better than Strategy 2.

Fig. 26 shows that the average target existence probability of the three strategies rapidly decreases in the initial stage, and there is a significant difference in the middle stage. In the 100 to 400 seconds, Strategy 3 performs better than Strategy 1 because in Expert System 2 (Table 3), it is set that when the target discovery ratio is greater than 80%, the weights of the optimization function will be adjusted, so that the search decision changes from probability first to uncertainty first. Under no communication constraints in Strategy 1, each UAV will get the discovered target number timely. Therefore, the expert system of Strategy 1 will adjust the objective function's weights earlier than Strategy 3, slowing down the average probability decline rate of Strategy 1.

As the search time increases, the three strategies' average uncertainty and average probability decrease rate are damping. This phenomenon is because the repeated search area will inevitably increase as the searched area increases. From the average results of multiple simulations, the search performance of Strategy 3 is significantly better than Strategy 2 and is close to Strategy 1.

**E. Evaluation of GA calculation time**

The calculation time of the genetic algorithm (GA) is mainly affected by the prediction sequence length ($m$) and the jump grid value ($j$). Based on the simulation environment of section VII A, shut down the expert system of the UAV. The computational time of GA is systematically analyzed by setting different prediction sequence lengths and jump grid values for simulation. A single core runs the simulation program in MATLAB, and the computer's CPU is i7-9700k. The statistical results of the average GA calculation time are shown in Fig. 27.

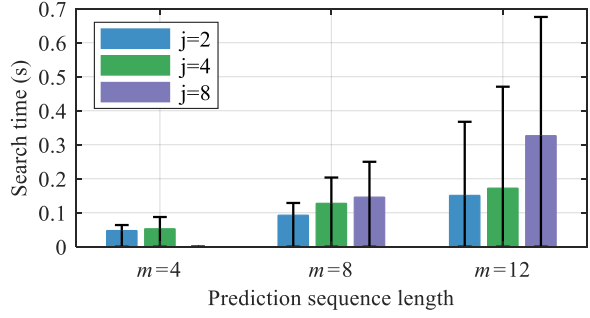

**Fig. 27. Statistical results of average GA calculation time**

Fig. 27 shows that the average calculation time of GA increases with the increase of $m$ and $j$. Moreover, the maximum calculation time of GA also has the same trend. According to (10) and (14), if $u_i(k)$ has $n$ possible values, The number of $\mathbf{U}_i(k)$ in the predictive action sequence set $\bar{\mathbf{U}}_i(k)$ is $n^m$. According to (11) and (12), the larger the space of $\bar{\mathbf{U}}_i(k)$, the longer the solution time of the optimization problem. Therefore, the average calculation time of GA will increase with the increase of $m$. From (8), the jump grid decision method reduces the possible value of $u_i(k)$ to $n = 3$, ensuring that the GA has an ideal solution speed. According to (15), the larger the $j$, the larger the turning radius of the UAV. When $j$ is large, the GA may be unable to find a feasible solution within a limited number of iterations, so it has to reduce $j$ through the maneuverability dynamic adjustment strategy (Fig. 11, in Section V B) and re-iterate. Therefore, the average calculation time of GA will increase with the increase of $j$.

In the case of the same $m$ and $j$, the difference in GA's calculation time is caused by environments. When the UAV is in an open area, the GA may reach the termination error of the objective function after 2 or 3 iterations. When the UAV is in an area with large environmental complexity, the GA makes it challenging to find the optimal solution, increasing the calculation time. In simulations, the calculation time of GA accounts for more than 98% of the decision time, and the maximum calculation time of GA is less than the decision interval (1s). Therefore, from the perspective of computational requirements, the approach proposed in this paper is practical in engineering applications.

**F. Cooperative search in dynamic environment**

A new simulation environment is designed to verify the universality of the proposed approach. Set up a heterogeneous cluster with five rotor UAVs and one fixed-wing UAV. The rotor UAV's field of view is a 40m × 60m rectangle and forward along the flight direction by 40m. The UAV 4 is set to exit the search task in 100s. The task area is 2000m × 800m, with seven targets and eight denied areas (obstacles). In order to increase the dynamic characteristics of the environment, denied areas are set to move within the task area at a speed of 4m/s, and the initial movement direction is random. Compared to the simulation in Section VII B, the current simulation environment has changed in task area, number of UAVs, UAV's field of view, number of targets, number of denied areas, and motion of denied areas. The simulation result is shown in Fig. 28.

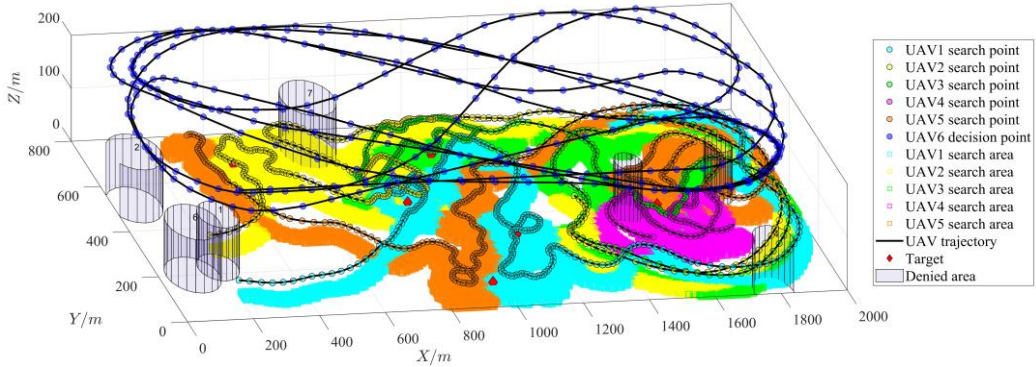

**Fig. 28. Cooperative search of heterogeneous UAVs in dynamic environment. More details can be found in the attached video.**

The simulation result shows that heterogeneous UAVs still have ideal search performance in complex and dynamic environments, indicating that the approach proposed in this paper has excellent environmental adaptability and universality.

**G. Discussion of the results**

In this section, a series of simulations verifies the cooperative search approach of heterogeneous UAVs. Under the communication constraints, the search performance of homogeneous and heterogeneous clusters is compared. Different search strategies' average target existence probability and environmental uncertainty are analyzed. The calculation time of GA is evaluated, and the effectiveness of the algorithm in a dynamic environment is tested. The discussion of the results is as follows:

(1) The jump grid decision method ensures the search decisions meet UAVs' maneuverability constraints and consider UAVs' flight speed and turning radius. Under the distributed MPC framework, the jump grid decision method can conveniently update search positions, making it suitable for solving the discrete optimization problem.

(2) The parameter dynamic selection method enables UAVs to quickly discover targets, explore unknown areas, and avoid threat areas, significantly improving the rationality of search decisions at different task stages.

(3) The search information transmission method enables UAVs to share historical search information under small bandwidth requirements and replicate the search states of other UAVs locally, significantly improving the information transmission efficiency under communication constraints.

(4) Using fixed-wing UAVs as communication nodes increases the communication frequency of UAVs in the cluster and improves the information flow between rotor UAVs. Simulation shows that under communication constraints, the cooperative search performance of a heterogeneous cluster is significantly improved compared to a homogeneous cluster.

(5) The cooperative search approach of heterogeneous UAVs proposed in this paper has great universality and can be applied to complex and dynamic task environments.

## VIII. Conclusion

This paper proposes a systematic approach for heterogeneous cluster cooperative search tasks. The work enables UAVs with different maneuverability to perform search tasks cooperatively. By setting up communication relay nodes, the difference in UAV's flight speed and altitude is fully played, and the information transmission efficiency between UAVs is improved. With the proposed methods, the UAVs can have excellent flight safety and decision-making adaptability and exhibit ideal cooperative search performance under communication constraints.

The feasibility of the proposed methods is verified through simulations in this work. In the future, we will implement the cooperative search approach in ROS and carry out software/hardware in the loop simulations in a Linux environment. We will equip onboard companion computers and flight controllers on rotor UAVs and fixed-wing UAVs. The companion computer runs cooperative search algorithms, and the flight controller with PX4 firmware controls the motion of the UAV. Optical pods will be mounted on rotor UAVs with target recognition and positioning software to detect targets. Further, we will conduct heterogeneous cluster flight tests to validate the methods proposed in this paper.

## IX. Data availability statement

Some or all data, models, or code generated or used during the study are available in a repository online in accordance with funder data retention policies. The program for UAVs cooperative search used in simulations can be downloaded at

https://github.com/LiuZhenchang/UAV_Cooperative_Search.git